\documentclass[draftcls,onecolumn,12pt]{IEEEtran1.6b}
\usepackage{amsmath}
\usepackage{amssymb}
\usepackage{times}
\usepackage{color}
\usepackage{graphicx}
\usepackage{cite}
\usepackage{pst-all}

\newcommand{\insertfig}[4]{
\begin{figure}[ht]
\centerline{\includegraphics[width=#1\columnwidth]{#2.eps}}
\caption{#3}\label{#4}\end{figure}}

\DeclareMathAlphabet{\mathsfbf}{OT1}{cmss}{sbc}{n}

\newcommand{\example}[2]{
\begin{center}
\parbox{0.9\columnwidth}{
\rule{0.9\columnwidth}{0.5mm}\\
\noindent {\bf Example~#1:} 
#2\\
\rule{0.9\columnwidth}{0.5mm}
}
\end{center}
}

\newcommand{\EE}{\mathbb{E}} 
\newcommand{\PP}{\mathbb{P}} 

\newcommand{\ee}{{\rm e}}
\newcommand{\ii}{{\rm i}}  
\newcommand{\dd}{{\rm\,d}} 

\newcommand{\av}{{\bf a}}
\newcommand{\bv}{{\bf b}}

\newcommand{\lv}{{\bf l}}

\newcommand{\pv}{{\bf p}}
\newcommand{\qv}{{\bf q}}

\newcommand{\tv}{{\bf t}}

\newcommand{\zerov}{{\bf 0}}

\newcommand{\Am}{{\bf A}}

\newcommand{\Fm}{{\bf F}}

\newcommand{\Id}{{\bf I}}

\newcommand{\Tm}{{\bf T}}
\newcommand{\Um}{{\bf U}}
\newcommand{\Wm}{{\bf W}}

\newcommand{\Zm}{{\bf Z}}

\newcommand{\Lc}{{\cal L}}

\newcommand{\Nc}{{\cal N}}

\newcommand{\Pc}{{\cal P}}
\newcommand{\Qc}{{\cal Q}}

\renewcommand{\Tc}{{\cal T}}  
\newcommand{\Uc}{{\cal U}}

\newcommand{\lambdav}{\hbox{\boldmath$\lambda$}}

\newcommand{\Lambdam}{\hbox{\boldmath$\Lambda$}}

\newcommand{\diag}{{\hbox{diag}}}

\def\trace{\mathsf{Tr}}

\def\ben{\begin{enumerate}}
\def\beq{\begin{equation}}
\def\beqa{\begin{eqnarray}}
\def\bit{\begin{itemize}}
\def\een{\end{enumerate}}
\def\eeq{\end{equation}}
\def\eeqa{\end{eqnarray}}
\def\eit{\end{itemize}}

\def\non{\nonumber\\}

\def\union{\mathop{\cup}\limits}
\def\average{\mathop{\EE}\limits}

\title{Bandlimited Field Reconstruction for\\ Wireless Sensor Networks}
\author{Alessandro Nordio \quad \and 
Carla-Fabiana Chiasserini \quad \and 
Emanuele Viterbo \\
Politecnico di Torino -- Dipartimento di Elettronica \\
C. Duca degli Abruzzi 24, I-10129 Torino (Italy)\\ 
e-mail: {\tt <name>@polito.it}
\thanks{This work was supported through the 
PATTERN project}
}

\begin{document}
\maketitle

\begin{abstract}
Wireless sensor networks are often used for environmental monitoring 
applications. In this context sampling and reconstruction of a physical 
field is one of the most important problems to solve. 
We focus on a bandlimited field and find under which conditions on the
network topology the reconstruction of the field is successful, with a
given probability. 
We review irregular sampling theory, and analyze the problem 
using random matrix theory.
We show that even a very irregular spatial distribution of sensors  may
lead to a successful signal reconstruction, provided that the number of
collected samples is large enough with respect to the field bandwidth.
Furthermore, we give the basis to analytically determine the probability of
successful field reconstruction.
\end{abstract}


\section{Introduction}
\label{sec:introduction}

One of the most popular applications of wireless sensor networks is 
environmental monitoring.
In general, a physical phenomenon (hereinafter also called sensor field or 
physical field) may vary over both space and time, with some 
band limitation in both domains.
In this work, we address the problem of sampling and reconstruction of a 
spatial field at a fixed time instant. We focus on a bandlimited 
field (e.g., pressure and temperature),
and assume that sensors are randomly deployed over a geographical 
area to sample the phenomenon of interest. 

Data are transfered from the sensors to a common data-collecting unit, 
the so-called sink node.
In this work, however, we are concerned only with the 
reconstruction of the sensor field,
and we do not address issues related to information transport.
Thus, we assume that all data is correctly
received at the sink node. Furthermore, we assume that the sensors have
a sufficiently high precision so that the quantization error is negligible,
and the sensors position is known at the sink node. 
The latter assumption implies that nodes are either located at 
pre-defined positions, or, if randomly deployed, their location 
can be acquired (see \cite{Hightower01,Hu04,Moore04} for
a description of node location methods in sensor networks).

Our objective is to investigate the relation between the network 
topology and the probability of successful reconstruction of the field 
of interest. The success of the reconstruction algorithm strongly
depends on the given machine precision, since it may fail to invert 
some ill-conditioned Toeplitz matrix (see Section~\ref{sec:irr_sampling}).

More specifically, we pose the following question: {\em under which 
conditions on the network 
topology (i.e., on the sample distribution) 
the sink node successfully reconstructs the signal with a given 
probability?} 
The solution to the problem seems to be hard to find, even under the 
simplifying assumptions we described above.  

The main contributions of our work are summarized below. 
\begin{enumerate}
\item[{\em (i)}] We first consider deterministic sensor locations.
By reviewing irregular sampling theory~\cite{Feichtinger95}, we show some 
sufficient conditions on the number of sensors to be deployed 
and on how they should be spatially  spaced so as to successfully 
reconstruct the measured field.
\item[{\em (ii)}] We then consider a random network topology and analyze the 
problem using random matrix theory. 
We identify the conditions under which the filed reconstruction is 
successful with a fixed probability, and we show that even a very 
irregular spatial distribution of sensors may lead to a successful 
signal reconstruction, provided that the number of collected samples 
is large enough with respect to the field bandwidth.
\item[{\em (iii)}] Finally we provide the theoretical basis to estimate the 
required number of active sensors, given the field bandwidth.
\end{enumerate}

\section{Related work}

Few papers have addressed the 
problem of sampling and reconstruction in sensor networks.
Efficient techniques for spatial sampling in sensor networks are 
proposed in~\cite{Perillo04,Willett04}. 
In particular~\cite{Perillo04} 
presents an algorithm to determine which sensor subsets should be 
selected to acquire data from an area of interest and which nodes should 
remain inactive to save energy.  
The algorithm chooses sensors in such a way that the node positions 
can be mapped into a blue noise binary pattern. 
In~\cite{Willett04}, an adaptive sampling is 
described, which allows the central data-collector to vary the 
number of active sensors, i.e., samples, according to the desired 
resolution level.
Data acquisition is also studied in~\cite{Kumar03}, 
where the authors consider a unidimensional field, uniformly sampled at 
the Nyquist frequency by low precision sensors. The authors show that 
the number of sensors (i.e., samples) can be traded-off with the 
precision of sensors.
The problem of the reconstruction of a bandlimited signal
from an irregular set of samples at unknown locations is addressed
in \cite{Marziliano00}. There, different solution methods are proposed, and 
the conditions for which there exist multiple solutions or a unique solution 
are discussed. 

Note that our work significantly differs from the studies above because 
we assume that the sensors location
are known (or can be determined \cite{Hightower01,Hu04,Moore04}) 
and the sensor precision is
sufficiently high so that the quantization error is negligible.
The question we pose is instead under which conditions (on the network system)
the reconstruction of a bandlimited signal is successful
with a given probability.


\section{Irregular sampling of band-limited signals}
\label{sec:irr_sampling}
Let us consider the one-dimensional model where $r$ sensors, 
located in the normalized interval $[0,1)$, measure the value
of a band-limited signal $p(t)$. 
As a first step, we assume that the position of the sensors 
sampling the field are deterministic and known, and the sensors can 
represent each sample with a sufficient number of
bits so that the quantization error is negligible.
Let $t_q\in [0,1)$ for $q=1 \ldots, r$
be the deterministic locations of the sampling points ordered increasingly
and $p(t_q)$ the corresponding samples. 

A strictly band-limited signal over the interval $[0,1)$ can be written as
the weighted sum of $M'$ harmonics in terms of Fourier series
\begin{equation} \label{blsig}
p(t) = \sum_{k=-{M'}}^{M'} a_k \ee^{2\pi \ii k t}
\end{equation}
Note that for real valued signals the Fourier coefficients satisfy 
the relation $a_k^*=a_{-k}$ and that the series (\ref{blsig})
can be represented as a sum of cosines.

The reconstruction problem can be formulated as follows:
\medskip

\noindent {\em given $r$ pairs $[t_q,p(t_q)]$ 
for $q=1,\ldots, r$ and $t_q \in [0,1)$ find the band-limited signal in (\ref{blsig}) 
uniquely specified by the sequence of its Fourier coefficients $a_k$}.

\medskip
Let the reconstructed signal be
\begin{equation} \label{blsighat}
\hat{p}(t) = \sum_{k=-M}^M  \hat{a}_k \ee^{2\pi \ii k t}
\end{equation}
where the  $\hat{a}_k$ are the corresponding Fourier coefficients
up to the $M$-th harmonic. In general, the reconstruction procedure will minimize
$\|p(t)-\hat{p}(t)\|^2$ if $M < M'$
and give $p(t)=\hat{p}(t)$ if $M = M'$.

Consider the $(2M+1)\times r$ matrix $\Fm$ whose $(k,q)$-th 
element is defined by
\begin{equation*}
(\Fm)_{k,q}=\frac{1}{\sqrt{r}}\ee^{2\pi \ii k t_q}~~~
\begin{array}{l} k= -M,\ldots, M \\ 
q= 1,\ldots,r\end{array}
\end{equation*}
the vector $\hat{\av}=[\hat{a}_{-M},\ldots,\hat{a}_0,\ldots,\hat{a}_M]^{\rm T}$ 
of size $2M+1$ and the vector\\ 
$\pv=[p(t_1),\ldots,p(t_r)]^{\rm T}$.
We have the following linear system~\cite{Feichtinger95}:
\begin{equation} \label{linsystem}
\Fm\Fm^\dagger \hat{\av} = \Fm \pv
\end{equation}
where $(\cdot)^\dagger$ is the conjugate transpose operator.
Let us denote $\Tm=\Fm\Fm^\dagger$ and $\bv = \Fm\pv$, hence (\ref{linsystem})
becomes $\Tm\hat{\av}=\bv$ and then $\hat{\av}=\Tm^{-1}\bv$.

When the samples are equally spaced in the interval $[0,1)$, i.e.,
$t_q=(q-1)/r$, we observe that the matrix $\Fm$ is a unitary matrix 
($\Fm\Fm^\dagger=\Tm=\Id_{2M+1}$)~\footnote{The symbol $\Id_n$ 
represents the $n$ by $n$ identity matrix} 
and its rows are orthonormal vectors of an inverse DFT matrix.
In this case (\ref{linsystem}) gives the first $M$ Fourier coefficients
of sample sequence $\pv$.

When the samples $t_q$ are not equally spaced, the matrix 
$\Fm$ is no longer unitary and the matrix $\Tm$ becomes a 
$(2M+1)\times (2M+1)$ Hermitian Toeplitz matrix
\[
\Tm =\Tm^\dagger = \left( \begin{array}{cccc}
r_0 & r_1 & \cdots & r_{2M} \\
r_{-1} & r_0 & \cdots & r_{2M-1} \\
       &     & \ddots &  \\
r_{-2M}&     & \cdots & r_0 
\end{array}\right)
\]
where
\begin{eqnarray}\label{Tm}
(\Tm)_{k,m} = r_{k-m} =
\frac{1}{r}\sum_{q=1}^{r}\ee^{2\pi \ii (k-m) t_q}~~~~~~~k,m=-M\ldots,M
\end{eqnarray}

The above Toeplitz matrix $\Tm$ is uniquely defined 
by the $4M+1$ variables
\begin{equation}
\label{eq:r_ell}
r_\ell=\frac{1}{r}\sum_{q=1}^{r}\ee^{2\pi \ii \ell t_q}~~~~~
\ell=-2M,\ldots 2M
\end{equation}

The solution of (\ref{linsystem}), which involves the inversion of 
$\Tm$, requires some care if the condition number of $\Tm$ 
(or equivalently of $\Fm$) becomes large.
We recall that the condition number of $\Tm$ is defined as
\begin{equation}
\kappa = \frac{\lambda_{\max}}{\lambda_{\min}}
\label{eq:condition_number}
\end{equation}
where $\lambda_{\max}$ and $\lambda_{\min}$ are the largest and 
the smallest eigenvalues of $\Tm$, respectively. The base-10 logarithm of 
$\kappa$ is an estimate of how many base-10 digits are lost in solving a linear system 
with that matrix.

In practice, matrix inversion is usually performed by algorithms 
which are very sensitive to small eigenvalues, especially 
when smaller than the machine precision.
For this reason in~\cite{Feichtinger95} a preconditioning technique 
is used to guarantee a bounded condition number when the maximum 
separation between consecutive sampling points is not too large. 
More precisely, by defining $w_q=(t_{q+1}-t_{q-1})/2$ for 
$q=1 \ldots, r$, where $t_0=t_r-1$
and $t_{r+1}=1+t_1$, and by letting $\Wm=\diag(w_1, \ldots, w_r)$,
the preconditioned system becomes
\[
\Tm_w\hat{\av} = \bv_w
\]
where $\Tm_w=\Fm\Wm\Fm^\dagger$ and $\bv_w = \Fm\Wm\pv$.
Let us define the maximum gap between consecutive sampling points as
\[
\delta = \max(t_q-t_{q-1}).
\]
In~\cite{Feichtinger95} it is shown that, when $\delta<1/2M$,we have: 
\begin{equation}
\kappa(\Tm_w) \leq \left(\frac{1+2\delta M}{1-2\delta M}\right)^2
\label{eq:Tmw}
\end{equation}
This result generalizes the Nyquist sampling theorem
to the case of irregular sampling, but only gives a {\em sufficient} 
condition for perfect reconstruction when the condition number
is compatible with the machine precision.
Unfortunately, when $\delta>1/2M$, the result (\ref{eq:Tmw}) does not hold.

In Figure~\ref{fig:signal6} and~\ref{fig:signal10} we present two examples 
of reconstructed signals from irregular sampling, using~(\ref{linsystem}).
Figure~\ref{fig:signal6} refers to the case $M=10$ and $r=26$, where the 
samples have been randomly selected over the interval $[0,0.8)$. 
The signal is perfectly reconstructed even if large 
gaps are present ($\delta>0.2$, i.e., $\delta>1/2M$). 
In Figure~\ref{fig:signal10}, $r=21$ samples of the same 
signal of Figure~\ref{fig:signal6} have been taken randomly over 
the entire window $[0,1)$. 
Due to the bad conditioning of the matrix $\Tm$ (i.e., very low 
eigenvalues), the algorithm fails in reconstructing the signal
due to machine precision underflow.

Driven by these observations, the objective of our work is to provide
conditions for the successful reconstruction of the sampled field, by
using a probabilistic approach. In the following we give 
a probabilistic description of the condition number, 
without explicitly considering preconditioning.

\section{The random matrix approach: unsuccessful signal reconstruction}
The above results are based on deterministic locations of the sampling points.
In this section we discuss instead the case where the sampling 
points $t_q$ are i.i.d. random variables with uniform distribution
${\Uc}[0,1)$. In other words we consider the case where the matrix $\Tm$ 
is random and completely defined by the random vector $\tv=[t_1,\ldots,t_r]$.
We introduce here the parameter $\beta$ as the ratio of the two-sided signal bandwidth $2M+1$
and the number of sensors $r$
\begin{equation}
\beta = \frac{2M+1}{r}.
\label{eq:beta}
\end{equation} 
In the following we consider the asymptotic case where the values of $M$
and $r$ grow to infinity while $\beta$ is kept constant. 
We then show that properties of systems with finite $M$ and $r$ are
well approximated by the asymptotic results.

We focus here on the expression of the probability of unsuccessful
signal reconstruction, i.e., the probability that the reconstruction algorithm 
fails given the machine precision $\epsilon$, the signal bandwidth $M$, and the number 
of sensors $r$.
For a given realization of $\Tm$ and for finite values of $M$ and $r$ we denote by 
$\lambdav=[\lambda_1,\ldots,\lambda_{2M+1}]$ the vector of eigenvalues, and by
$\lambda_{\rm min} = \min(\lambdav)$ and $\lambda_{\rm max} = \max(\lambdav)$ 
the minimum and maximum eigenvalues, respectively.
Also let $f_{M,\beta}(x)$ be the empirical probability density function (pdf)
of the eigenvalues of $\Tm$ for a finite $M$ and $\beta$ 
and let $f_{\beta}(x)$ be the limiting eigenvalue pdf in the asymptotic case
(i.e., when $M$ and $r$ grow to infinity with constant $\beta$) \cite{Tulino}.
The random variable $\lambda_{\rm min} = \min(\lambdav)$,
and the condition number $\kappa$ 
have pdf $f_{M,\beta}^{\rm min}(x)$ and $f_{M,\beta}^{\kappa}(x)$, respectively.
The corresponding cumulative density functions (cdf) are denoted by $F_{M,\beta}(x)$, $F_{\beta}(x)$,
$F_{M,\beta}^{\rm min}(x)$, and $F_{M,\beta}^{\kappa}(x)$.

\subsection{Some properties of the eigenvalue distribution}
\label{sec:eig_distr}
We first analyze by Montecarlo simulation some properties
of the distribution $f_{M,\beta}(x)$. Figure~\ref{fig:pdf_b025} shows
histograms of $f_{M,\beta}(x)$ for $M=1,4,10,90$, $\beta=0.25$, and bin width of $0.1$.
Notice that, as $M$ increases with constant $\beta$, the histograms of $f_{M,\beta}(x)$ 
seem to converge to $f_{\beta}(x)$, only depending on $\beta$.
Indeed, looking at the figure, one can notice that the difference between the 
curves for $M=10$ and $M=90$ is negligible.
Although we report in Figure~\ref{fig:pdf_b025} only the case for $\beta=0.25$, 
we observed the same behavior for any value of $\beta$. 
We therefore conclude that $M=10$ is large enough to provide a good 
approximation of $f_{\beta}(x)$.

In Figure~\ref{fig:pdf_all_beta} we show histograms of $f_{M,\beta}(x)$ for $\beta=0.15,0.25,0.35,0.45,0.55$ 
and values of $M$ around $100$.
For $\beta$ larger than $0.35$ the distribution shows oscillations and tends to infinity while
$x$ approaching $0$. On the other hand, for $\beta$ lower than $0.35$ the pdf does not oscillate
and tends to $0$ while $x$ approaching $0$. In order to better understand this behavior for small $x$, 
which can be heavily affected by the bin width, in Figure~\ref{fig:cdf} we consider the cdf
$F_{M,\beta}(x)$ in the log-log scale, for various values of $\beta$ ranging from $0.1$ to $0.8$ and $M=200$.
The dashed curves represent the simulated cdf. Surprisingly they show a linear behavior for small 
values of $x$ and for any value of $\beta$. 
This is evidenced by the solid lines which are the tangents to
the dashed curves at $F_{M,\beta}(x)=10^{-2}$.
The slope of the lines is parameterized by $\beta$.
In our simulations the machine precision is approximately $\epsilon = 10^{-16}$ and, hence, 
values of $x<\epsilon$ cannot be represented since they are treated as zero by the algorithm. 
Indeed the simulated pdfs loose their linear behavior while approaching $x=\epsilon$ 
(see  the case $\beta=0.8$ in Figure~\ref{fig:cdf}).
We conclude that for $x \ll 1$ the cdf $F_{\beta}(x)$ can be approximated by
\begin{equation}
F_{\beta}(x) \approx b x^a
\label{eq:Fx}
\end{equation}
where $a=a(\beta)$ and $b=b(\beta)$ are both functions of $\beta$.
By deriving (\ref{eq:Fx}) with respect to $x$ we obtain the approximate expression 
for the pdf:
\begin{equation}
f_{\beta}(x) \approx a(\beta) b(\beta) x^{a(\beta)-1}
\label{eq:fx}
\end{equation}
From (\ref{eq:fx}) it can be seen that the function $a(\beta)$ represents the slope of $F_{\beta}(x)$ 
in the log-log scale for $x \ll 1$. 
Note that in order $x^{a(\beta)-1}$ to be integrable in $[0,c)$, for any positive constant $c$, 
the condition $a(\beta)>0$ should be satisfied. Note also from Figure~\ref{fig:cdf} that the slope 
$a(\beta)=1$ is obtained for $\beta \approx0.35$. For this value of $\beta$ the approximate pdf is
constant for $x \ll 1$, which is consistent with the results in Figure~\ref{fig:pdf_all_beta}.

Some additional considerations can be drawn from Figure~\ref{fig:pdf_loglog}, which presents 
the pdf of $f_{M,\beta}(x)$ for $\beta=0.25,0.50,0.75$ and $M=200$.
It is interesting to note that for any value of $\beta$, large eigenvalues are less likely to appear
than very small eigenvalues. This is evident by observing that for $x \gg 1$ the pdf falls 
to $-\infty$ much faster than for $x \ll 1$. This consideration is of great relevance when 
discussing the condition number distribution.   

\subsection{Distribution of the minimum eigenvalue}
For finite $M$ the cdf of $\lambda_{\rm min}$ can be computed as follows
\begin{eqnarray*}
F_{M,\beta}^{\rm min}(x) &=& \PP(\lambda_{\rm min}<x | M)  \\
                               &=& \PP(\min(\lambdav)<x | M)
\end{eqnarray*}
In general the random variables $\lambda_1,\ldots,\lambda_{2M+1}$ are not independent.
However, considering sufficiently large values of $M$ (namely, $M \geq 10$), 
we can write the following upper bound 
for $F_{M,\beta}^{\rm min}(x)$:
\begin{equation}
F_{M,\beta}^{\rm min}(x) \leq (2M+1) F_{\beta}(x).
\label{eq:approx1}
\end{equation}
This is obtained by assuming that the eigenvalues are independent
with pdf equal to the limiting eigenvalue distribution.
The simulation results presented in Figure~\ref{fig:failure} confirm the
expression in (\ref{eq:approx1}).
The figure shows the cdfs of $\lambda$ and $\lambda_{\rm min}$ in the log-log scale
for $\beta=0.25,0.50,0.75$ and $M=40$.
The cdf of $\lambda_{\rm min}$ also shows a linear behavior for $x \ll 1$.
In the log-log scale, according to (\ref{eq:approx1}), the two cdfs should be separated 
by $\log_{10}(2M+1)$. 
In our case: $M=40$ and $\log_{10}(2M+1) \approx 1.91$.
As is evident from the figure, this upper bound is extremely 
tight, especially for low values of $\beta$.

\subsection{Distribution of the condition number}
Here we describe the condition number distribution. The condition number is defined by~(\ref{eq:condition_number}).
As noted at the end of Section~\ref{sec:eig_distr} the minimum eigenvalue dominates 
the ratio $\lambda_{\rm max}/\lambda_{\rm min}$.
This fact is more evident in Figure~\ref{fig:failure_kn}, where we compare 
the distributions of the condition number and of the minimum eigenvalue,
for $\beta=0.25$ and $M=10,20,40$.
The three dashed curves on the left represent the pdf of the minimum eigenvalue. 
The solid lines on the right represent the pdf of the condition number for the same values of $M$.
The two set of distributions look very similar.
We define $y=\log_{10}x$, $\gamma^{\rm min}_{M,\beta}(y) = \log_{10} f^{\rm min}_{M,\beta}(10^y)$ and
$\gamma^{\kappa}_{M,\beta}(y) = \log_{10} f^{\kappa}_{M,\beta}(10^y)$.
By observing the results in Figure~\ref{fig:failure_kn}, the following relation holds:
\[
\gamma^{\kappa}_{M,\beta}(y) \approx \gamma^{\rm min}_{M,\beta}(-y + d)
\]
where $d$ is a parameter. In the plot, for each value of $M$ the circles represent the
above approximation where the parameter $d$ is set to $1/3$.
The same considerations hold for any value of $\beta$.
Converting the above approximation into the linear scale, we obtain:
\[
f^{\kappa}_{M,\beta}(x) \approx f^{\rm min}_{M,\beta}\left(\frac{10^d}{x}\right)
\]
and by taking the derivative of both sides of (\ref{eq:approx1}) with respect to $x$, we finally obtain
\[
f^{\kappa}_{M,\beta}(x) \approx (2M+1)f_{\beta}\left(\frac{10^d}{x}\right)
\]
which holds for $x \gg 1$.

\subsection{Summary}
In this section we have given numerical evidence of the following facts: 
\begin{itemize}
\item the condition number distribution is dominated by
the distribution of the minimum eigenvalue of $\Tm$; 
\item the distribution of the minimum eigenvalue is upper bounded by a simple function of
the asymptotic distribution of the eigenvalues of $\Tm$.
\end{itemize}
Thus, in the following we focus on $f_{\beta}(x)$; indeed,
knowing $f_{\beta}(x)$ we could obtain the probability that the minimum
eigenvalue is below a certain threshold, i.e., that the condition number is 
less the machine precision.

\section{Some analytic results on the eigenvalue pdf}
We now derive some analytic results on the asymptotic eigenvalue distribution, $f_{\beta}(x)$.
Ideally we would like to analytically compute $f_{\beta}(x)$, however such a 
calculation seems to be prohibitive. 
Therefore, as a first step we compute the closed form expression of the moments of the asymptotic 
eigenvalue distribution, $\EE[\lambda^p]$. Note that, 
if all moments are available, the an analytic expression of $f_{\beta}(x)$ can be derived through 
its moment generating function, by applying the inverse Laplace transform.

In the limit for $M$ and $r$ growing to infinity with constant $\beta$ 
the expression of $\EE[\lambda^p]$ can be easily obtained from the powers of $\Tm$.
Indeed $\Tm$ is an Hermitian matrix and can be decomposed as $\Tm=\Um\Lambdam\Um^\dagger$,
where $\Lambdam=\diag(\lambdav)$ is a diagonal matrix containing the eigenvalues 
of $\Tm$ and $\Um$ is the matrix of eigenvectors. It follows that
\begin{eqnarray}
\trace\{\Tm^p\} &=& \trace\left\{\left(\Um\Lambdam\Um^\dagger\right)^p\right\}  \non
                &=& \trace\{\Um\Lambdam^p\Um^\dagger\} \non
                &=& \trace\{\Um^\dagger\Um\Lambdam^p\} \non
                &=& \trace\{\Lambdam^p\} \non
                &=& \sum_{i=1}^{2M+1} \lambda_i^p
\end{eqnarray}
Then:
\begin{eqnarray}
  \lim_{\substack{M,r\rightarrow +\infty \\\frac{2M+1}{r}=\beta}}\frac{1}{2M+1} \trace\{\EE\left[\Tm^p\right]\} 
&=&   \lim_{\substack{M,r\rightarrow +\infty \\\frac{2M+1}{r}=\beta}}\frac{1}{2M+1} \EE\left[\sum_{i=0}^{2M}\lambda_i^p\right] \non
&=&   \EE\left[\lim_{\substack{M,r\rightarrow +\infty \\\frac{2M+1}{r}=\beta}}\frac{1}{2M+1} \sum_{i=0}^{2M}\lambda_i^p\right] \non
&=&  \EE\left[\lambda^p\right]
\label{eq:lambda_p}
\end{eqnarray}
Please notice that since $\Tm$ is a Toeplitz matrix the Grenander-Szeg\"o~\cite{Grenander58} 
theorem could be employed in the limit for $M\rightarrow +\infty$.
Unfortunately in this case the theorem is not applicable since 
all entries of $\Tm$ depend on the matrix size $M$. 

From (\ref{eq:lambda_p}) and (\ref{eq:r_ell}) we obtain:
\begin{equation}
\EE[\lambda^p] = \lim_{\substack{M,r\rightarrow +\infty \\\frac{2M+1}{r}=\beta}}
\frac{1}{(2M+1)r^p}~\sum_{\qv \in \Qc}~\sum_{\lv \in \Lc}~\average_{\tv}\left[\exp\left(2\pi \ii \sum_{i=1}^p 
t_{q_i}(\ell_i-\ell_{[i+1]})\right)\right]
\label{eq:lambda_p2}
\end{equation}
where
\begin{eqnarray*}
\Qc &=& \left\{\qv~|~\qv=[q_1, \ldots, q_p],~~q_i=1,\ldots,r \right\} \\
\Lc &=& \left\{\lv~|~\lv=[\ell_1, \ldots, \ell_p],~~\ell_i=0,\ldots,2M \right\}
\end{eqnarray*}
and where the sign $[\cdot]$ refers to the modulo $p$ operator\footnote{For simplicity 
here we follow the convention $[p] = p$ and $[p+1] = 1$.}.
The average is performed over the random vector $\tv=[t_1,\ldots, t_r]$.

Let now $\Pc$ be the set of integers from 1 to $p$
\begin{equation} 
\Pc=\{1,\ldots,p\}.
\label{eq:P_set} 
\end{equation}
Let $\qv \in \Qc$ and let $1\le k(\qv) \le p$ be the number of distinct values assumed by 
the entries of $\qv$. Such values can be arranged, in order of appearance, in the vector
$\hat{\qv}=[\hat{q}_1, \ldots, \hat{q}_{k(\qv)}]$ where the entries $\hat{q}_j$ are all distinct.
Using $\qv$ and $\hat{\qv}$ we create the subsets $\Pc_1(\qv),\ldots, \Pc_{k(\qv)}(\qv)$
of $\Pc$ defined by
\begin{equation}
\Pc_j(\qv)= \left\{ i\in \Pc~|~q_i=\hat{q}_j \right\}.
\end{equation}
Such subsets are non-empty and disjoint 
($\Pc_j \neq \emptyset$, $\union_j \Pc_j = \Pc$, and $\Pc_j \cap \Pc_h = \emptyset$ for $j \neq h$).
Finally we define $\tau(\qv)$
\[\tau(\qv)=\left\{ \Pc_1(\qv), \ldots, \Pc_{k(\qv)}(\qv)\right\}\] 
as the partition of $\Pc$ induced by $\qv$.

\medskip
\example{1}{
Let $p=6$ and $\qv=[4,9,5,5,4,3]$. 
Then, by (\ref{eq:P_set}), $\Pc=\{1,2,3,4,5,6\}$. We have $k(\qv) = 4$
distinct values which we arrange, in order of appearance, in the vector $\hat{\qv}=[4,9,5,3]$. Then
\[
\begin{array}{ll} 
\Pc_1(\qv)= \{1,5\} & (q_1=q_5=\hat{q}_1),\\
\Pc_2(\qv)= \{2\}   & (q_2=\hat{q}_2),\\
\Pc_3(\qv)= \{3,4\} & (q_3=q_4=\hat{q}_3),\\ 
\Pc_4(\qv)= \{6\}   & (q_6=\hat{q}_4),
\end{array}
\]
and $\tau(\qv)=\{\{1,5\},\{2\},\{3,4\},\{6\}\}$.}
\medskip

For any given $\qv\in \Qc$, using the definition of $\Pc_j(\qv)$, we notice that the argument 
of the average operator in (\ref{eq:lambda_p2}) factorizes in $k(\qv)$ parts, i.e.
\[ \exp\left(2\pi \ii \sum_{i=1}^p t_{q_i}(\ell_i-\ell_{[i+1]})\right) 
= \prod_{j=1}^{k(\qv)} \exp\left(2\pi \ii  t_{\hat{q}_j} 
\sum_{i \in \Pc_j(\qv)}\ell_i-\ell_{[i+1]} \right)
\]
each depending on a single random variable $t_{\hat{q}_j}$. 
Then from (\ref{eq:lambda_p2}) we have:
\begin{eqnarray}
\EE[\lambda^p] &=& \lim_{\substack{M,r\rightarrow +\infty \\\frac{2M+1}{r}=\beta}} 
\frac{1}{(2M+1)r^p}~\sum_{\qv \in \Qc}~\sum_{\lv \in \Lc}~\average_{\tv}\left[ \prod_{j=1}^{k(\qv)} 
\exp\left(2\pi \ii  t_{\hat{q}_j} \sum_{i \in \Pc_j(\qv)}\ell_i-\ell_{[i+1]} \right)\right] \non
&=&  \lim_{\substack{M,r\rightarrow +\infty \\\frac{2M+1}{r}=\beta}} 
\frac{1}{(2M+1)r^p}~\sum_{\qv \in \Qc}~\sum_{\lv \in \Lc}~\prod_{j=1}^{k(\qv)}
~\average_{t_{\hat{q}_j}} \left[\exp\left(2\pi \ii t_{\hat{q}_j}
\sum_{i \in \Pc_j(\qv)}\ell_i-\ell_{[i+1]} \right)\right] \non
&=&  \lim_{\substack{M,r\rightarrow +\infty \\ \frac{2M+1}{r}=\beta}}
\frac{1}{(2M+1)r^p}~\sum_{\qv \in \Qc}~\sum_{\lv \in \Lc}~\prod_{j=1}^{k(\qv)} 
\delta\left( \sum_{i \in \Pc_j(\qv)}\ell_i-\ell_{[i+1]} \right)
\label{eq:lambda_p3}
\end{eqnarray}
where $\delta(\cdot)$ is the Kronecker's delta.
Expression (\ref{eq:lambda_p3}) can be further simplified by observing that
\begin{itemize}
\item there exist $r(r-1)\cdots(r-k+1) = r!/(r-k)!$ vectors $\qv\in \Qc$ generating a certain given 
partition of $\Pc$ made of $k$ subsets,
\item for a given $\qv$ the expression
\begin{eqnarray}
\zeta_{2M}(\qv)=\sum_{\lv \in \Lc}~\prod_{j=1}^{k(\qv)}
~\delta\left( \sum_{i \in \Pc_j(\qv)}\ell_i-\ell_{[i+1]} \right)
\label{eq:zeta}
\end{eqnarray}
is a polynomial in the variable $2M$, since it represents
the number of points with integer coordinates contained in the
hypercube $[0,\ldots,2M]^p$ and satisfying the $k(\qv)$ constraints
\begin{equation} 
\sum_{i \in \Pc_j(\qv)}\ell_i-\ell_{[i+1]}=0
\label{eq:constraints}
\end{equation}
We show in Appendix~\ref{app:constraints} that one of these constraints is always redundant
and that the number of linearly independent constraints is exactly $k(\qv)-1$.
By consequence the polynomial $\zeta_{2M}(\qv)$ has degree $p-k(\qv)+1$.
\end{itemize}
Let $\Tc_p$ be the set of distinct partitions of $\Pc$ generated by all vectors $\qv\in \Qc$, then
from (\ref{eq:lambda_p3}) we obtain:
\begin{eqnarray}
\EE[\lambda^p] &=&\lim_{\substack{M,r\rightarrow +\infty \\ \frac{2M+1}{r}=\beta}}
\frac{1}{(2M+1)r^p}~\sum_{\qv \in \Qc}~\sum_{\lv \in \Lc}~\prod_{j=1}^{k(\qv)} 
\delta\left( \sum_{i \in \Pc_j(\qv)}\ell_i-\ell_{[i+1]} \right) \non
 &\overset{(a)}{=}& \lim_{\substack{M,r\rightarrow +\infty\\ \frac{2M+1}{r}=\beta}} 
\frac{1}{(2M+1)r^p} \sum_{\tau \in \Tc_p}  \sum_{\qv \Rightarrow \tau} \zeta_{2M}(\qv) \non
 &\overset{(b)}{=}& \lim_{\substack{M,r\rightarrow +\infty\\ \frac{2M+1}{r}=\beta}} 
\frac{1}{(2M+1)r^p} \sum_{\tau \in \Tc_p}  \frac{r!}{(r-k(\tau))!} \zeta_{2M}(\tau)
\label{eq:lambda_p3a}
\end{eqnarray}
where 
\begin{itemize}
\item the notation $\sum_{\qv \Rightarrow \tau}$ represents
the sum over all vectors $\qv$ generating a certain given partition $\tau$,
\item the equality $(a)$ has been obtained by substituting (\ref{eq:zeta}), and
\item the equality $(b)$ holds because the number of vectors $\qv$ generating a
given partition $\tau$ is $r!/(r-k(\tau))!$.
\end{itemize}
We point out that the functions $k(\qv)$ and $\zeta_{2M}(\qv)$ 
depend only on the partition $\tau(\qv)$ induced by $\qv$.
Since in the third line of~(\ref{eq:lambda_p3a}) we removed the dependence
on the vectors $\qv$, the expression of $\EE[\lambda^p]$ is now function of the partitions $\tau$ only. 
Then with a little abuse of notation, in the following
we refer to the functions $k$ and $\zeta_{2M}$ as $k(\tau)$ and $\zeta_{2M}(\tau)$, respectively.

Taking the limit we finally obtain:
\begin{eqnarray}
 \EE[\lambda^p] &=&  \sum_{\tau \in \Tc_p} v(\tau)\beta^{p-k(\tau)} \non
 &=& \sum_{k=1}^p \left(\sum_{\tau \in \Tc_{p,k}} v(\tau) \right) \beta^{p-k}
\label{eq:lambda_p4}
\end{eqnarray}
where $\Tc_{p,k}$ is the subset of $\Tc_p$ only containing partitions of size $k$,
and
\[ v(\tau) = \lim_{M \rightarrow +\infty} \frac{\zeta_{2M}(\tau)}{(2M)^{p-k+1}} \]
i.e. $v(\tau)$ is the coefficient%
\footnote{Notice also that the coefficient $v(\tau)$ represents
the volume of the {\em convex polytope} described by the constraints (\ref{eq:constraints})
when the variables $\ell_i$ are considered real and limited to the interval $[0,1]$. 
By consequence $0 \le v(\tau) \le 1$.}
of degree $(2M)^{p-k+1}$ of the polynomial $\zeta_{2M}(\tau)$.
Since $1\le k\le p$ from (\ref{eq:lambda_p4}) we note that $\EE[\lambda^p]$ is a polynomial 
in $\beta$ of degree $\beta^{p-1}$. Again, for the sake of clarity we give an example: 

\medskip
\example{2}{Let $p=6$ and $\qv$ given by Example 1.
The partition is $\tau=\{ \{1,5\}, \{2\}, \{3,4\}, \{6\}\}$.
Then the set of $k(\tau)=4$ constraints~(\ref{eq:constraints}) are given by:
\begin{eqnarray}
\ell_1 + \ell_5 &=& \ell_2 + \ell_6 \non
\ell_2 &=& \ell_3 \non
\ell_3 + \ell_4 &=& \ell_4+ \ell_5 \non
\ell_6 &=& \ell_1 \nonumber
\end{eqnarray}
The last equation is redundant since can be obtained summing up the first three constraints.
Simplifying we obtain $\ell_1=\ell_6$, and $\ell_2=\ell_3=\ell_5$.
Since each variable $\ell_i$ ranges from $0$ to $2M$, the number of integer solutions
satisfying the constraints is exactly $\zeta_{2M}(\tau)=(2M+1)^3$, and then $v(\tau)=1$.} 
\medskip

To compute (\ref{eq:lambda_p4}) we need to enumerate the partitions $\tau \in \Tc_p$.
First of all we notice that $\Tc_p$ represents the set of partitions of a $p$-element set
and thus has cardinality $|\Tc_p| = B(p)$ where $B(p)$ is the $p$-th {\em Bell number} 
or {\em exponential number}~\cite{bellnumbers},
and that the subset $\Tc_{p,k}$ has cardinality $S_{p,k}$ which is a 
{\em Stirling number of the second kind}~\cite{stirling2numbers}. 
An effective way to enumerate such partitions is to build a tree of depth $p$ as in 
Figure~\ref{fig:tree4}. A label is given to each node, starting from the root which is 
labeled by ``a''. The rule for building the tree is as follows: each node $\Nc$ generates $m+1$ leaves,
labeled in increasing order starting from ``a'', and $m$ is the number of distinct 
labels in the path from the root to the node $\Nc$.
The number of leaves of such a tree of depth $p$ is given by $B(p)$. Each path from the
root to a leaf represents a partition $\tau$ of the set $\Pc$.
For a given partition (or path in the tree) the subset $\Pc_j$ is the  
set of integers corresponding to the depths of the $j$-th label in the path.

\medskip
\example{3}{Let us consider $p=4$ and the path $[a,b,a,a]$ (see Figure~\ref{fig:tree4}).
In the path there are two distinct labels, namely ``a'' and ``b''; then $k(\tau)=2$. The label ``a'' 
is found at depths 1,3, and 4, while the label ``b'' is at depth 2.
The partition of $\Pc=\{1,2,3,4\}$ is then given by 
$\tau=\{ \{1,3,4\}, \{2\}\}$. This partition (or path) contributes to the expression of 
$\EE[\lambda^p]=\EE[\lambda^4]$ with the term $v(\tau)\beta^{p-k} = \beta^2$ 
since in this case $v(\tau)=1$.
}
\medskip

Using the procedure described above we can derive in closed form any moment of $\lambda$.
Here we report the first few moments:
\begin{eqnarray*}
 \EE[\lambda]   &=& 1 \\
 \EE[\lambda^2] &=& 1+\beta \\
 \EE[\lambda^3] &=& 1+3\beta+\beta^2 \\
 \EE[\lambda^4] &=& 1+6\beta+\frac{20}{3}\beta^2 +\beta^3 \\
 \EE[\lambda^5] &=& 1+10\beta+\frac{70}{3}\beta^2 +\frac{40}{3}\beta^3 +\beta^4
\end{eqnarray*}
In practice the algorithm complexity prevents us from computing moments of order greater than $p=12$.
To the best of our knowledge, a closed form expression of the generic moment of $\lambda$ is still unknown.
If all moments were available, then an analytic expression of $f_{\beta}(x)$ could be derived through 
its moment generating function $\Psi_{\beta}(s)$ 
\begin{equation}
\Psi_{\beta}(s) = \int_0^{+\infty} f_{\beta}(x) \ee^{sx} \dd x 
             = \sum_{p=0}^{+\infty} \frac{\EE[\lambda^p]}{p!} s^p
\end{equation}
by applying the inverse Laplace transform.

\subsection{Validation}
We compare the moments of $\lambda$ obtained by simulation with those obtained with the above 
closed form analysis.
Table \ref{tab:moments} compares the exact values of the moments of $f_{\beta}(x)$,
and the values obtained by Montecarlo simulation,
for $\beta=0.25,0.50,0.75$ and $p=1,\ldots,5$. 
For each value of $\beta$ the Table shows three columns.
The first column, labeled ``Sim'' presents the values obtained by simulation,
using $M=200$. The second column, labeled ``Exact'',
reports the values obtained using (\ref{eq:lambda_p3})
{\em without} taking the limit (i.e., using finite values of $M$ and $r$).
The third column, labeled ``Limit'', presents the limit values obtained through 
(\ref{eq:lambda_p4}). 
The excellent match between simulation analytic results shows the validity of our findings.
\begin{table}
\caption{Comparison of the moments of $\lambda$ obtained by simulation 
and by closed form analysis for $M=200$, and $\beta=0.25,0.50,0.75$.}
\begin{center}
\begin{tabular}{|l||l|l|l||l|l|l||l|l|l|} \hline
&\multicolumn{3}{|c||}{$\beta=0.25$} &\multicolumn{3}{|c||}{$\beta=0.50$} &\multicolumn{3}{|c|}{$\beta=0.75$}\\ 
 & Sim & Exact & Limit & Sim & Exact & Limit & Sim & Exact & Limit\\ \hline
p=1 & 1.000 & 1.000 & 1.000 & 1.000  & 1.000  & 1.000 & 1.000 & 1.000   & 1.000  \\ \hline
p=2 & 1.249 & 1.249 & 1.250 & 1.499  & 1.499  & 1.500 & 1.748 & 1.748   & 1.750  \\ \hline
p=3 & 1.810 & 1.810 & 1.812 & 2.746  & 2.744  & 2.750 & 3.802 & 3.801   & 3.812  \\ \hline
p=4 & 2.926 & 2.925 & 2.932 & 5.778  & 5.771  & 5.792 & 9.630 & 9.620   & 9.672  \\ \hline
p=5 & 5.152 & 5.152 & 5.176 & 13.51  & 13.49  & 13.56 & 27.41 & 27.35   & 27.57 \\ \hline
\end{tabular}
\end{center}
\label{tab:moments}
\end{table}

\section{Conclusions}
\label{sec:conclusions}
We considered a large-scale wireless sensor network sampling
a physical field, and we investigated the 
relationship between the network topology and the 
probability of successful field reconstruction. 
In the case of deterministic sensor locations, we derived some
sufficient conditions for successful reconstruction, by reviewing
the literature on irregular sampling.
Then, we considered random network topologies, and 
employed random matrix theory. By doing so, we were able to
derive some conditions under which the field can be successfully
reconstructed with a given probability. 

A great deal of work still has to be done. However, to the best of our 
knowledge, this work is the first attempt at
solving the problem of identifying the conditions on random
network topologies for the reconstruction of sensor fields.
Furthermore, we believe 
that the basis we provided for an analytical study of the problem can be
of some utility in other fields besides sensor networks.

\appendices

\section{The constraints}
\label{app:constraints}
Let us consider a vector of integers $\qv$ of size $p$ 
partitioning the set $\Pc=\{1,\ldots,p\}$ in $k$ subsets
$\Pc_j$, $1\le j \le k$ and the set of $k$ constraints
\begin{equation}
\sum_{i \in \Pc_j}\ell_i-\ell_{[i+1]}=0.
\end{equation}
We first show that one of such constraint is always redundant.

\subsection{Redundant constraint}
Choose an integer $j$, $1\le j \le k$. Summing up together the constraints, except the
$j$-th, we get
\begin{eqnarray}
0 &=& \sum_{\substack{h=1\\ h\neq j}}^k \sum_{i \in \Pc_h}\ell_i-\ell_{[i+1]}\non
  &=& \sum_{i \in \Pc/\Pc_j} \ell_i-\ell_{[i+1]}\non
  &=& \sum_{i \in \Pc} \ell_i-\ell_{[i+1]} -  \sum_{i \in \Pc_j} \ell_i-\ell_{[i+1]} \non
  &=& - \sum_{i \in \Pc_j} \ell_i-\ell_{[i+1]}
\end{eqnarray}
which gives the $j$-th constraint
\[ \sum_{i \in \Pc_j} \ell_i-\ell_{[i+1]}=0.\]
Thus one of the constraints (\ref{eq:constraints}) is always redundant.
We now show that the remaining $k-1$ constraints are linearly independent.

\subsection{Linear independence}
The $k$ constraints (\ref{eq:constraints}), after some simplifications, can be rearranged 
in the form
\[ \Am\lv^{\rm T} =  \zerov \]
where $\Am$ is a $k\times p$ matrix and $\lv=[\ell_1,\ldots,\ell_p]$.
We have previously shown that the rank of $\Am$ is such that 
\begin{equation}
\rho(\Am)\le k-1
\label{eq:rank1}
\end{equation}
since one constraint is redundant and $k\le p$. We prove now that
the rank of $\Am$ is exactly $k-1$.

It is possible to write $\Am$ as $\Am = \Am' - \Am''$ where
$(\Am')_{ji}=1$ if $i\in \Pc_j$, and $0$ elsewhere.
The matrix $\Am'$ has rank $k$ since its rows are linearly independent
due to the fact that subsets $\Pc_j$ have empty intersection. 
Similarly $(\Am'')_{ji}=1$ if $[i-1]\in \Pc_j$, and $0$ elsewhere.
In practice the matrix $\Am''$ is the matrix $\Am'$ circularly shifted 
by one position to the right.
Hence it can be written as
\[ \Am'' = \Am'\Zm \]
where $\Zm$ is the $p\times p$ {\em right-shift matrix}, i.e. the entries of the 
$i$-th row of $\Zm$ are zeroes except for a ``1'' at position $[i+1]$.
By consequence 
\[ \Am = \Am' - \Am'\Zm = \Am'(\Id_p - \Zm), \]
where
\[ (\Id_p - \Zm) = \left[\begin{array}{ccccc}
+1     & -1     & 0      & \cdots & 0 \\
0      & \ddots & \ddots & \ddots & \vdots \\
\vdots & \ddots &        & \ddots & 0  \\
0      &        & \ddots & \ddots & -1 \\
-1     &  0     & \cdots &   0    & +1
\end{array}\right]\]
has rank $\rho(\Id_p - \Zm)=p-1$.
By consequence, using the property 
\begin{eqnarray}
\rho(\Am) &=  &  \rho(\Am'(\Id_p - \Zm)) \non
          &\ge&  \rho(\Am')+\rho(\Id_p - \Zm) -p \non
          & = &  k-1
\label{eq:rank2}
\end{eqnarray}
Considering together (\ref{eq:rank1}) and (\ref{eq:rank2}) we conclude
$\rho(\Am) = k-1$.

\newpage
\insertfig{0.7}{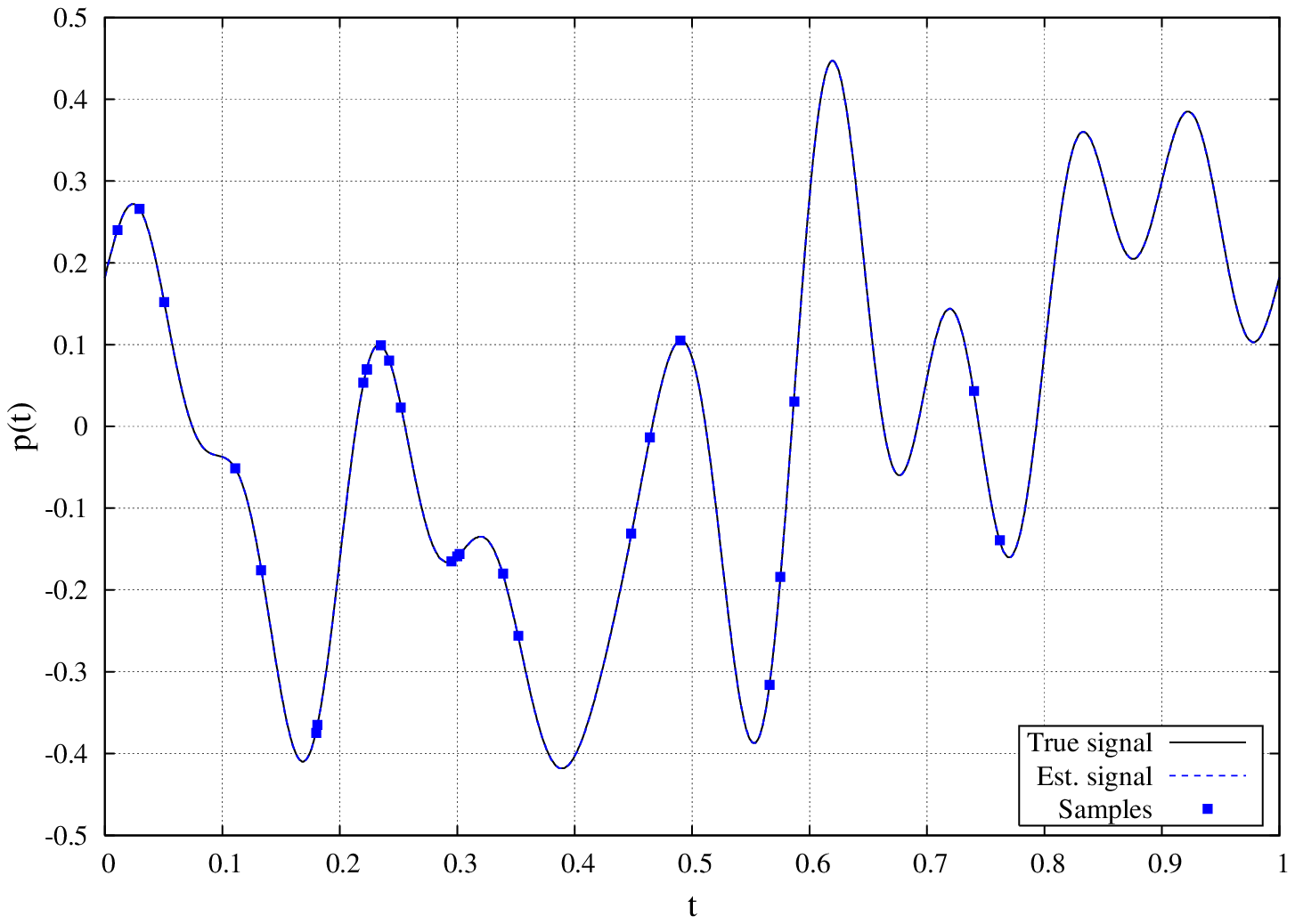}{Example of a reconstructed signal from 
irregular sampling, for $r=26$, $M=10$, $\beta=0.807$}{fig:signal6}
\insertfig{0.7}{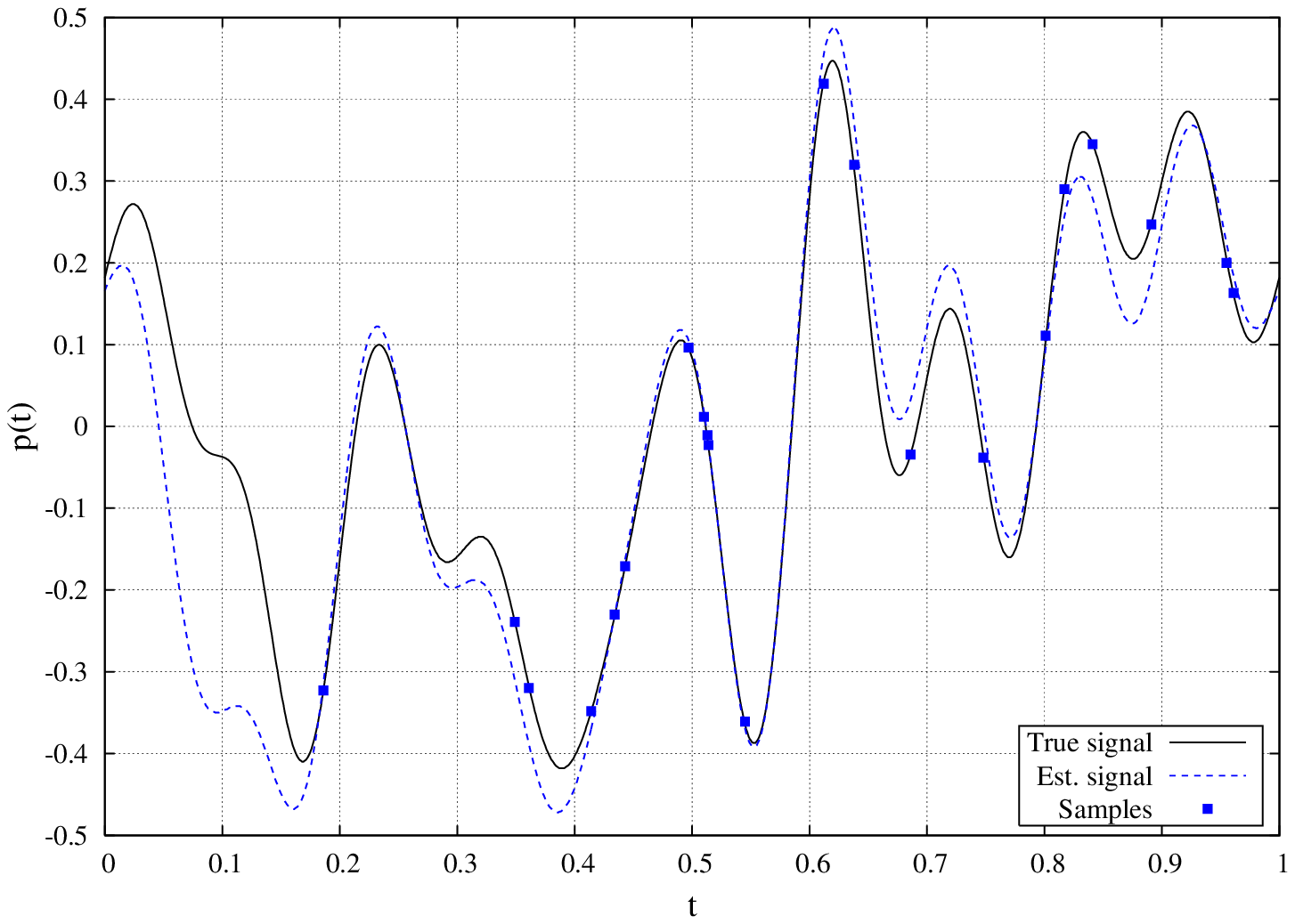}{Example of a badly reconstructed signal 
due to numerical instability for $r=21$, $M=10$, $\beta=1$}{fig:signal10}
\insertfig{0.7}{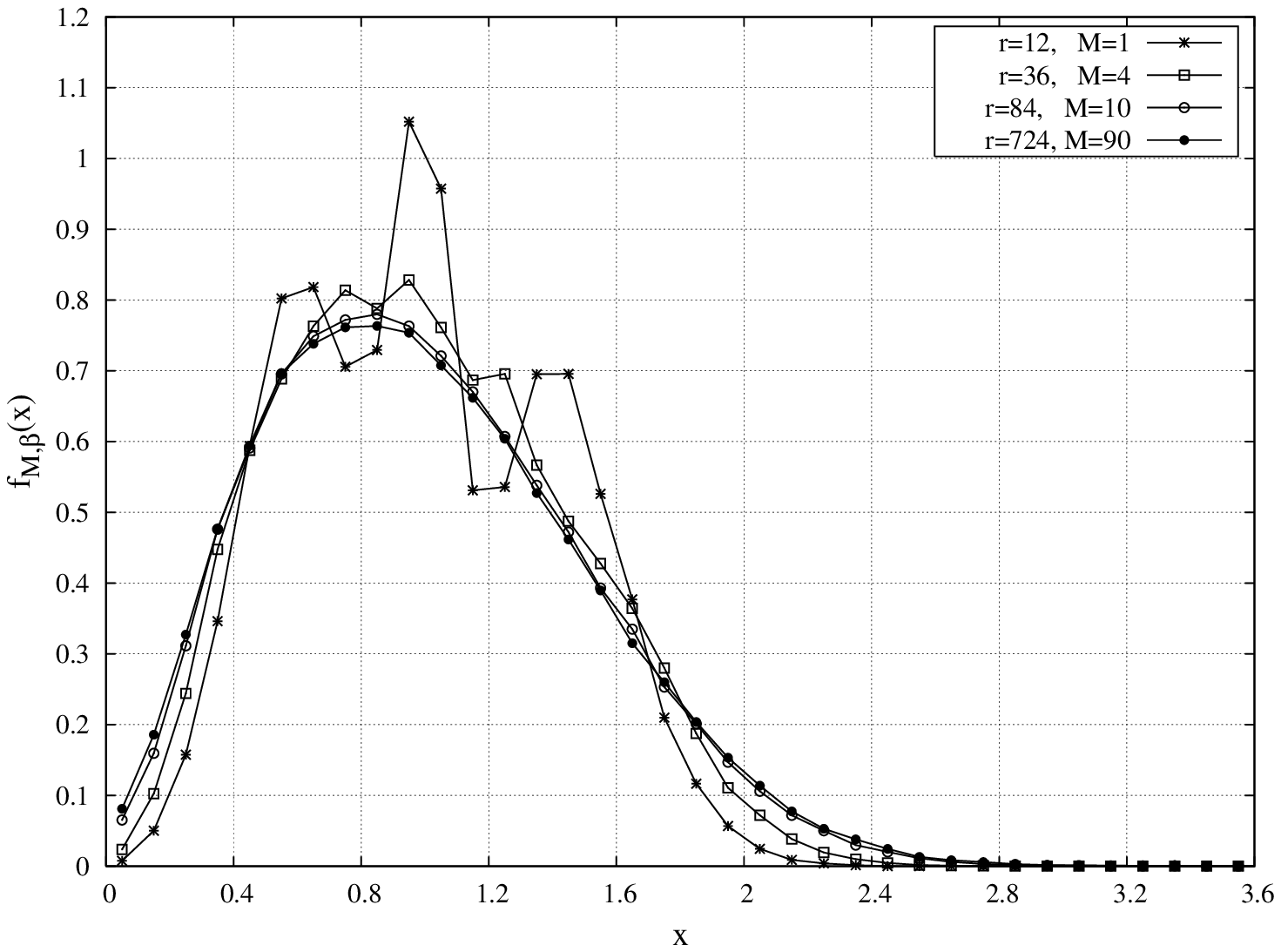}{Histograms of $f_{M,\beta}(x)$ for $\beta=0.25$ and 
increasing values of $M$}{fig:pdf_b025}
\insertfig{0.7}{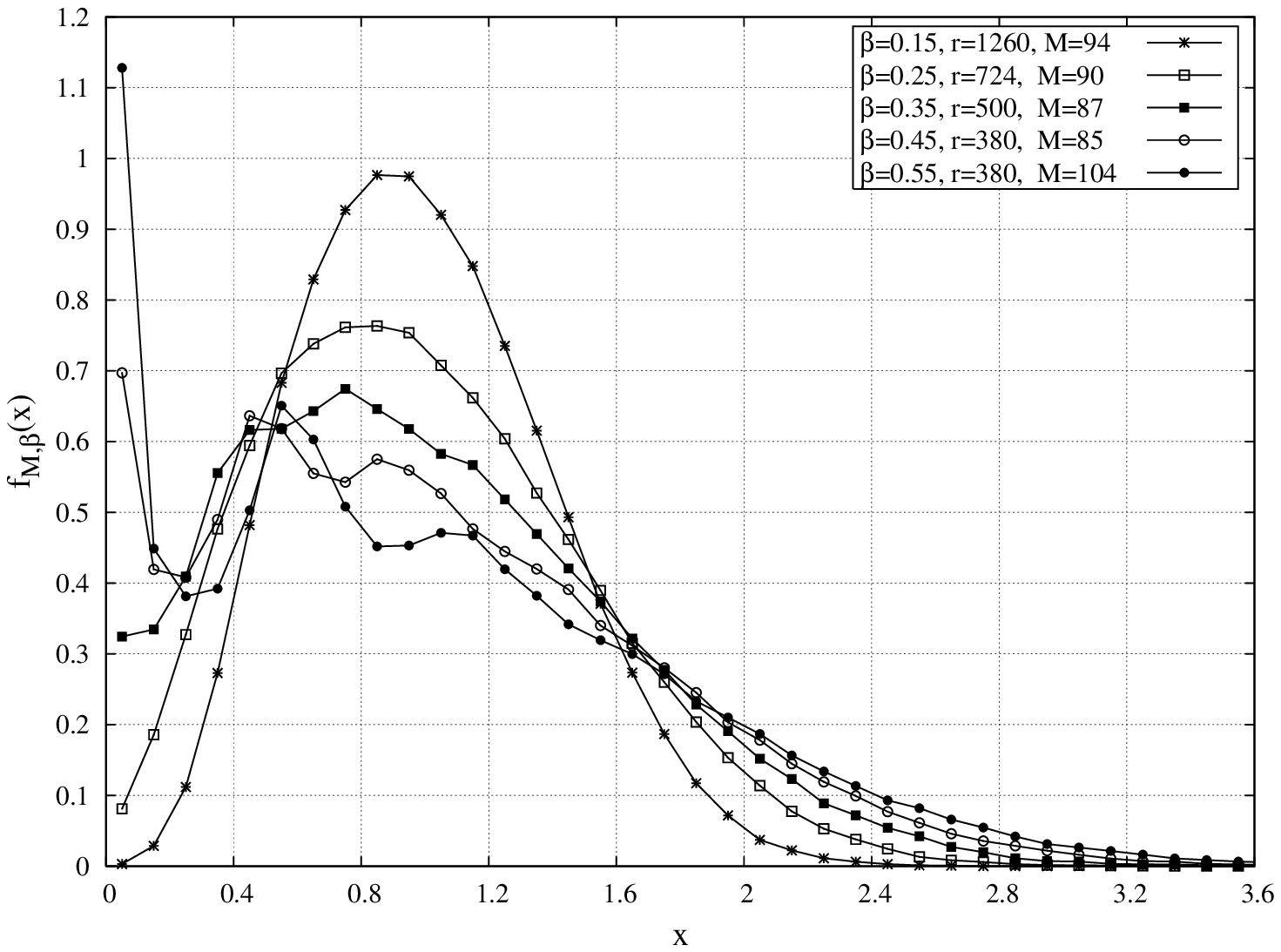}{Histograms of $f_{M,\beta}(x)$ for  
$\beta=0.15,0.25,0.35,0.45,0.55$}{fig:pdf_all_beta}
\insertfig{0.7}{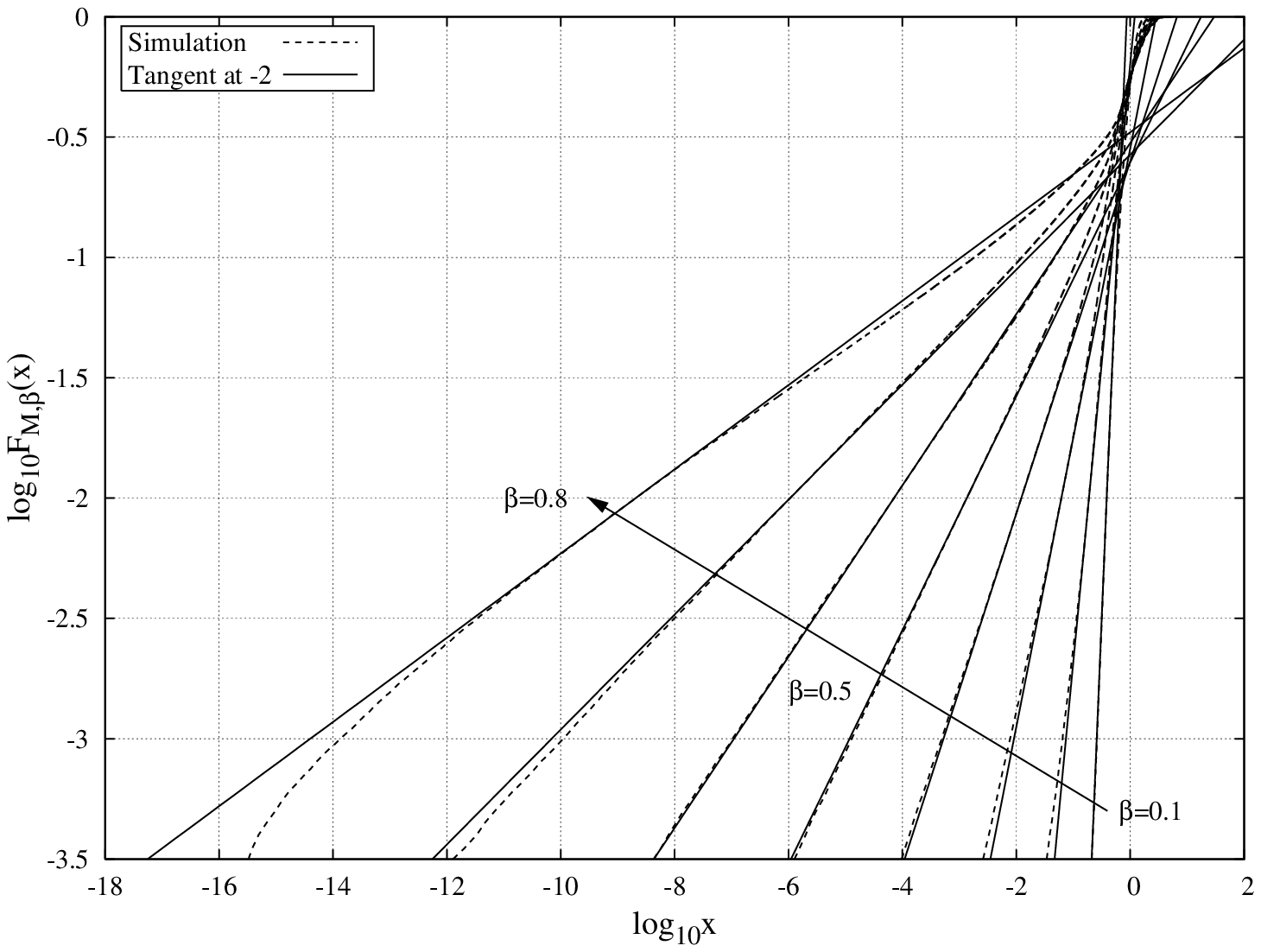}{Cumulative density function of $F_{M,\beta}(x)$ in the log-log scale 
for some values of $\beta$}{fig:cdf}
\insertfig{0.7}{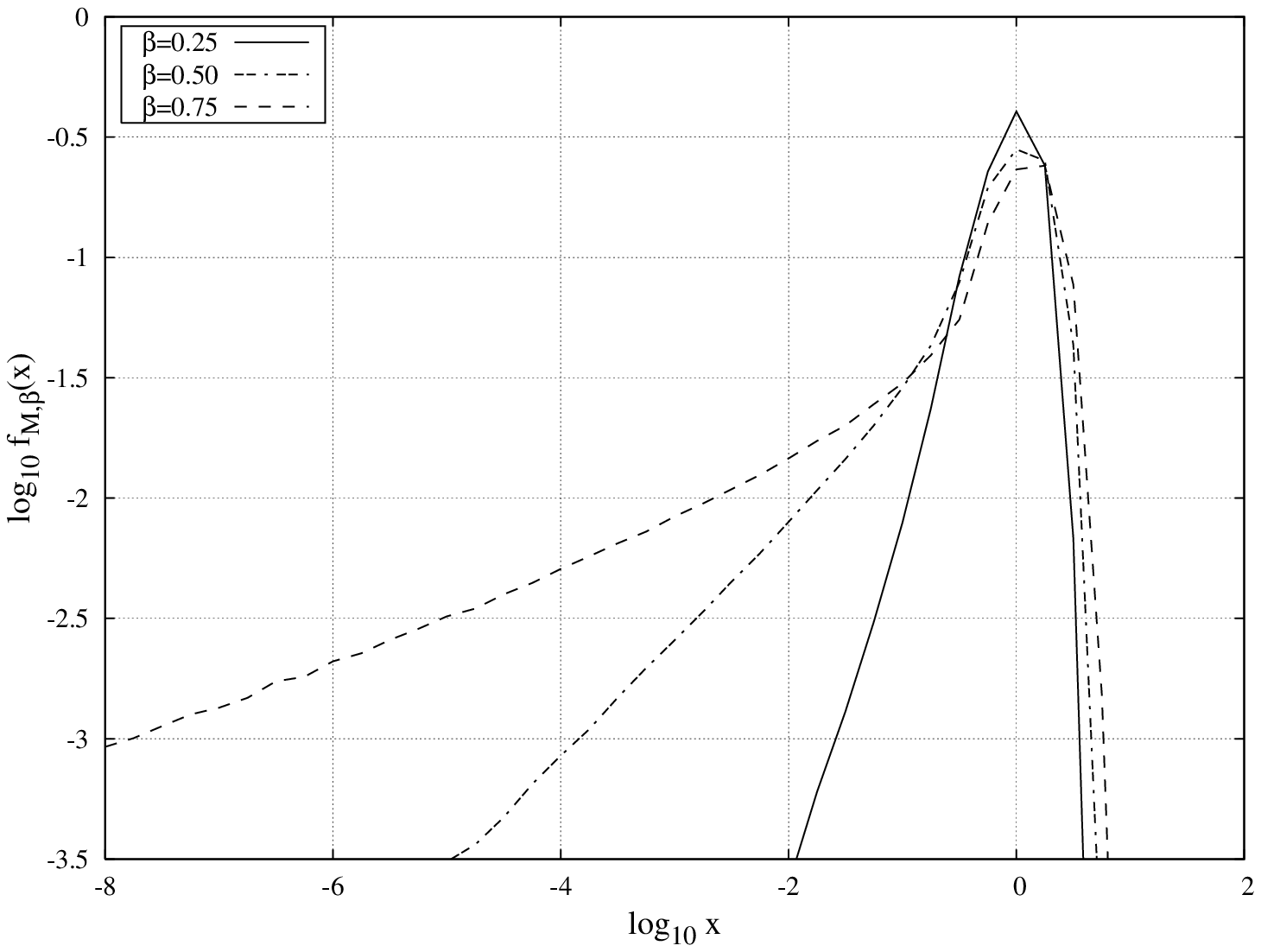}{Histograms of $f_{M,\beta}(x)$ in the log-log scale for 
$\beta=0.25,0.50,0.75$ and $M=200$}{fig:pdf_loglog}
\insertfig{0.7}{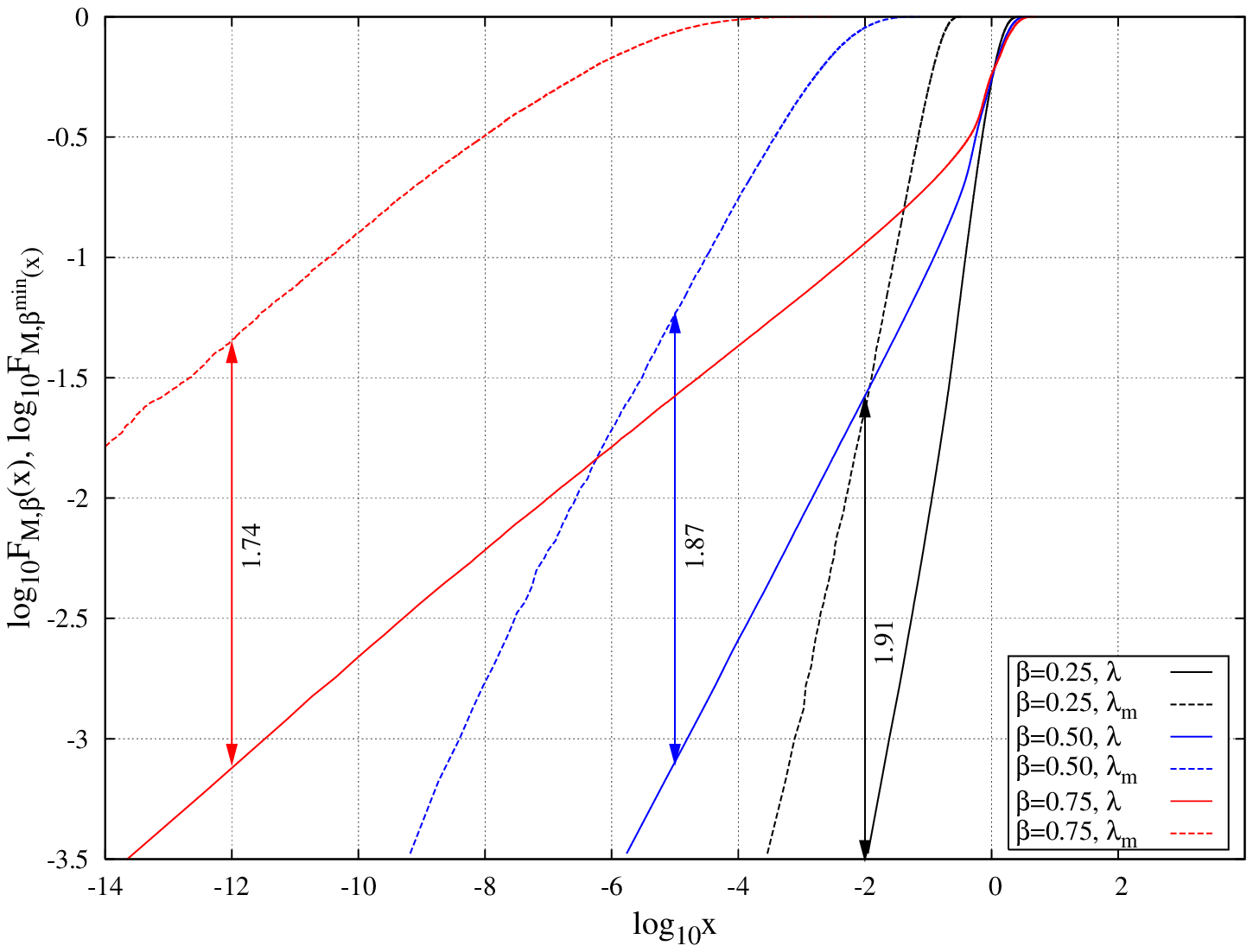}{Cumulative density functions $F_{M,\beta}(x)$ and 
$F_{M,\beta}^{\rm min}(x)$ in the log-log scale for $\beta=0.25,0.50,0.75$ 
and $M=40$}{fig:failure}
\insertfig{0.7}{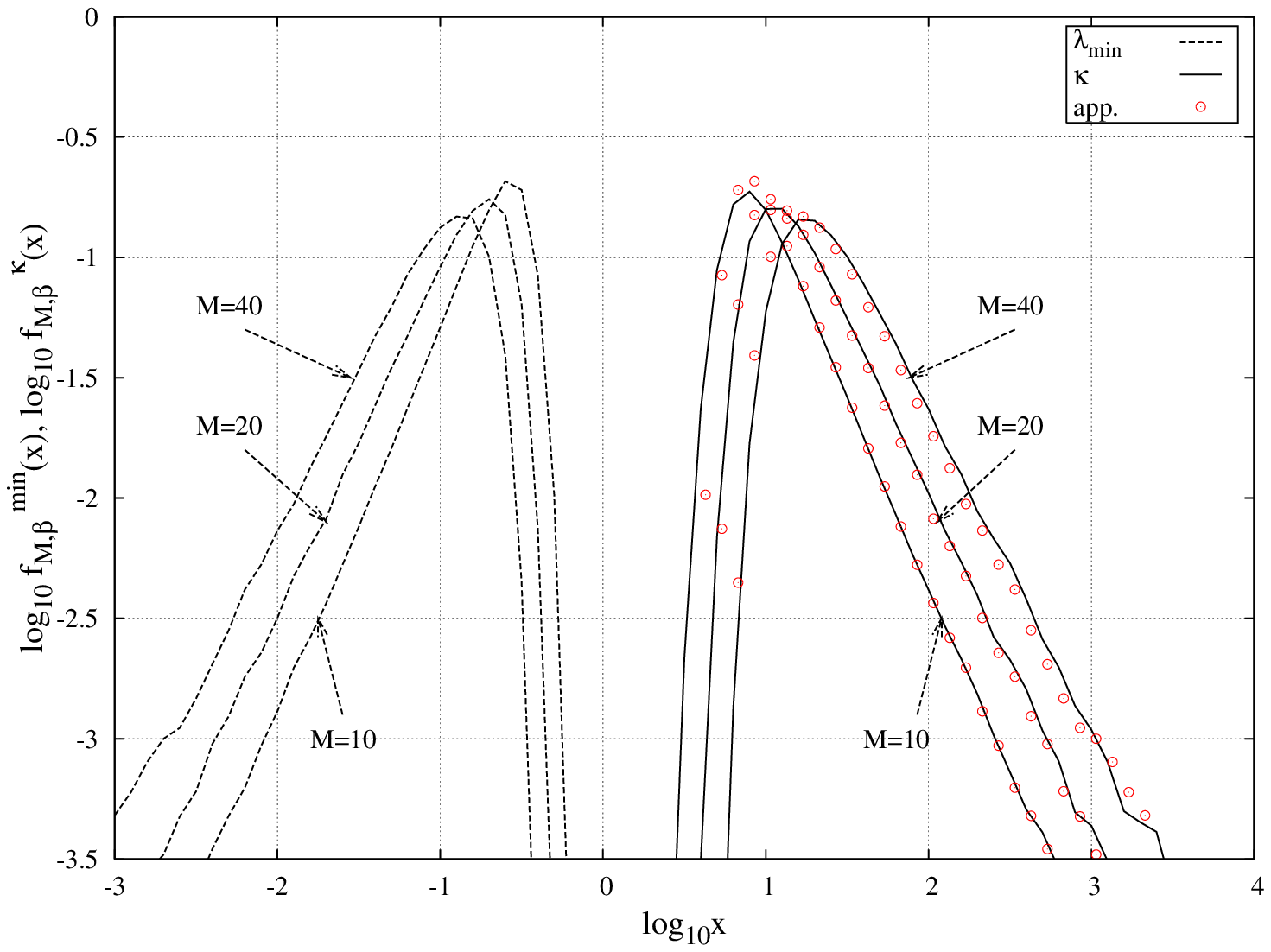}{Histograms of $f_{M,\beta}^{\rm min}(x)$ 
and $f_{M,\beta}^{\kappa}(x)$ in the log-log scale for $\beta=0.25$ 
and $M=10,20,40$}{fig:failure_kn}
\begin{figure}[h]
\centerline{\resizebox{0.9\columnwidth}{!}{
\input tree4.pstex_t}}
\caption{Partitions tree}
\label{fig:tree4}
\end{figure}

\end{document}